\documentclass[modern]{aastex62}

\usepackage{xspace}
\usepackage{amsmath}
\usepackage{newunicodechar,graphicx}
\usepackage{longtable}

\DeclareRobustCommand{\okina}{%
  \raisebox{\dimexpr\fontcharht\font`A-\height}{%
    \scalebox{0.8}{`}%
  }%
}
\newunicodechar{ʻ}{\okina}

\newcommand{\numax}{\mbox{$\nu_{\rm max}$}\xspace}
\newcommand{\Dnu}{\mbox{$\Delta \nu$}\xspace}

\newcommand{\muHz}{\mbox{$\mu$Hz}\xspace}

\newcommand{\rsun}{\mbox{$\mathrm{R}_{\sun}$}\xspace}

\newcommand\kepler{\emph{Kepler}\,}
\newcommand\ktwo{\emph{K2}\,}

\definecolor{linkcolor}{rgb}{0.1216,0.4667,0.7059}

\usepackage{xcolor, fontawesome}
\definecolor{twitterblue}{RGB}{64,153,255}
\newcommand\twitter[1]{\href{https://twitter.com/#1 }{\textcolor{twitterblue}{\faTwitter}\,\tt \textcolor{twitterblue}{@#1}}}

\graphicspath{{./}{figures/}}

\received{January 1, 2019}
\revised{January 7, 2019}
\accepted{\today}
\submitjournal{ApJ}

\shorttitle{The K2 Bright Star Survey}
\shortauthors{B. J. S. Pope et al.}

\begin{document}

\title{The K2 Bright Star Survey I: Methodology and Data Release}

\correspondingauthor{Benjamin J. S. Pope \twitter{fringetracker}}
\email{benjamin.pope@nyu.edu}

\author[0000-0003-2595-9114]{Benjamin J. S. Pope}
\affiliation{Center for Cosmology and Particle Physics, Department of Physics, New York University, 726 Broadway, New York, NY 10003, USA}
\affiliation{Center for Data Science, New York University, 60 Fifth Ave, New York, NY 10011, USA}
\affiliation{NASA Sagan Fellow}

\author[0000-0002-6980-3392]{Timothy R. White}
\affiliation{Sydney Institute for Astronomy, School of Physics A28, The University of Sydney, NSW 2006, Australia}
\affiliation{Stellar Astrophysics Centre, Department of Physics and Astronomy, Aarhus University, DK-8000 Aarhus C, Denmark}
\affiliation{Research School of Astronomy and Astrophysics, Mount Stromlo Observatory, The Australian National University, Canberra, ACT 2611, Australia}

\author[0000-0003-1540-8562]{Will M. Farr}
\affiliation{Center for Computational Astrophysics, Flatiron Institute, 162 Fifth Ave, New York, NY 10010, USA}
\affiliation{Department of Physics and Astronomy, Stony Brook University, Stony Brook, NY 11794, USA}

\author[0000-0002-0007-6211]{Jie Yu}
\affiliation{Max Planck Institute for Solar System Research, Justus-von-Liebig-Weg 3, 37077 G\"{o}ttingen, Germany}

\author[0000-0002-0371-1647]{Michael Greklek-McKeon}
\affiliation{Institute for Astronomy, University of Hawai\okina i, 2680 Woodlawn Drive, Honolulu, HI 96822, USA}

\author[0000-0001-8832-4488 ]{Daniel Huber}
\affiliation{Institute for Astronomy, University of Hawai\okina i, 2680 Woodlawn Drive, Honolulu, HI 96822, USA}

\author[0000-0003-1822-7126]{Conny Aerts}
\affiliation{Instituut voor Sterrenkunde, KU Leuven, Celestijnenlaan 200D, B-3001 Leuven, Belgium}
\affiliation{Department of Astrophysics, IMAPP, Radboud University Nijmegen, P.O. Box 9010, NL-6500 GL Nijmegen, The Netherlands}
\affiliation{Max-Planck-Institut f\"{u}r Astronomie, K\"{o}nigstuhl 17, D-69117 Heidelberg}

\author[0000-0003-1453-0574]{Suzanne Aigrain}
\affiliation{Oxford Astrophysics, Denys Wilkinson Building, University of Oxford, OX1 3RH, Oxford, UK}

\author[0000-0001-5222-4661]{Timothy R. Bedding}
\affiliation{Sydney Institute for Astronomy, School of Physics A28, The University of Sydney, NSW 2006, Australia}
\affiliation{Stellar Astrophysics Centre, Department of Physics and Astronomy, Aarhus University, DK-8000 Aarhus C, Denmark}

\author[0000-0001-9879-9313]{Tabetha Boyajian}
\affiliation{Department of Physics and Astronomy, Louisiana State University, 202 Nicholsom Hall, Baton Rouge, LA 70803, USA}

\author[0000-0003-1853-6631]{Orlagh L. Creevey}
\affiliation{Universit\'{e} C\^{o}te d'Azur, Observatoire de la C\^{o}te d'Azur, CNRS, Laboratoire Lagrange, Bd de l'Observatoire, CS 34229, 06304 Nice cedex 4, France}

\author[0000-0003-2866-9403]{David W. Hogg}
\affiliation{Center for Cosmology and Particle Physics, Department of Physics, New York University, 726 Broadway, New York, NY 10003, USA}
\affiliation{Center for Data Science, New York University, 60 Fifth Ave, New York, NY 10011, USA}
\affiliation{Center for Computational Astrophysics, Flatiron Institute, 162 Fifth Ave, New York, NY 10010, USA}
\affiliation{Max-Planck-Institut f\"{u}r Astronomie, K\"{o}nigstuhl 17, D-69117 Heidelberg}



\begin{abstract}
While the \kepler\ Mission was designed to look at tens of thousands of faint stars ($V \gtrsim 12$), brighter stars that saturated the detector are important because they can be and have been observed very accurately by other instruments. By analyzing the unsaturated scattered-light `halo' around these stars, we have retrieved precise light curves of most of the brightest stars in \ktwo fields from Campaign~4 onwards. The halo method does not depend on the detailed cause and form of systematics, and we show that it is effective at extracting light curves from both normal and saturated stars. The key methodology is to optimize the weights of a linear combination of pixel time series with respect to an objective function. We test a range of such objective functions, finding that \textit{lagged Total Variation}, a generalization of Total Variation, performs well on both saturated and unsaturated \ktwo targets. Applying this to the bright stars across the \ktwo Campaigns reveals stellar variability ubiquitously, including effects of stellar pulsation, rotation, and binarity. We describe our pipeline and present a catalogue of the 161 bright stars, with classifications of their variability, asteroseismic parameters for red giants with well-measured solar-like oscillations, and remarks on interesting objects. These light curves are publicly available as a High Level Science Product from the Mikulski Archive for Space Telescopes (MAST). \href{https://github.com/benjaminpope/k2halo}{\color{linkcolor}\faGithub} 
\end{abstract}


\section{Introduction} 
\label{sec:intro}

The \kepler Space Telescope was launched with a main goal of determining the frequency of Earth-sized planets around Solar-like stars \citep{2010sci...327..977b}. In order to explore these populations it was necessary to observe hundreds of thousands of stars, with the consequence that the \kepler exposure time and gain were set to optimally observe eleventh or twelfth-magnitude stars, while bright stars are saturated and intentionally avoided. In the two-wheeled revival as the \ktwo mission, the \kepler telescope observed a sequence of ecliptic-plane fields containing many more very-saturated stars \citep{2014PASP..126..398H}. While it is difficult to obtain precise light curves of these stars because of their saturation, they are some of the most valuable targets to follow up with photon-hungry methods such as interferometry and high-resolution spectroscopy, and they typically have long histories of previous observations. Dedicated bright-star space photometry missions such as MOST \citep{most} and the BRITE-Constellation \citep{brite,brite2} use very small telescopes (15\,cm and~3\,cm apertures respectively), to assemble time-series photometry of bright stars, but larger telescopes such as \kepler (0.95\,m) lead to higher-precision light curves.

The \kepler detector saturates at a magnitude of $K_p \sim 11.3$ in both long- (30\,min) and short (1\,min)-cadence data, since these both represent sums of 6\,s exposures \citep{Gilliland2010}. For objects brighter than this, excess electrons `bleed' into adjacent pixels in both directions along the column containing the star. Simple aperture photometry (SAP) -- adding all the flux contained in a window around the bleed column -- has recovered light curves with precisions close to the photon noise limit. Examples treated in the nominal {\it Kepler\/} mission are the prototype classical radial pulsator RR\,Lyr \citep[$V=7.2$;][]{Kolenberg2011}, the solar-like pulsators 16\,Cyg\,AB \citep[$V\approx 6$;][]{2012ApJ...748L..10M,2013MNRAS.433.1262W,2015ApJ...811L..37M} and $\theta\,$Cyg \citep[$V=4.48$;][]{Guzik2016}, and the massive eclipsing binary V380\,Cyg \citep[$V=5.68$;][]{tkachenko2014}. In the nominal \kepler mission SAP was only attempted for a few bright stars, and in \ktwo, the larger-amplitude spacecraft motion significantly increased the size of the required apertures for SAP photometry of very saturated stars, while also making their instrumental systematics more difficult to deal with. While the second-version pixel-level-decorrelation (PLD) pipeline EVEREST~2.0 was able to correct systematics in saturated SAP photometry \citep{everest2}, this is not possible for the very brightest stars whose bleed columns may run to the edge of the detector. Furthermore, bandwidth constraints meant that pixel data were not downloaded for many bright targets in \ktwo. 

In order to recover precise light curves of the brightest stars in \ktwo, we have therefore developed two main approaches, `smear' and `halo' photometry. Smear photometry \citep{Pope2016,smearcampaign} uses collateral `smear' calibration data to obtain a 1-D spatial profile with $\sim 1/1000$ of the flux on each CCD. This can be processed to recover light curves of stars that were not necessarily conventionally targeted and downloaded with active pixels, because smear data are recorded for all columns. The main disadvantage of this method is that it confuses all stars in the same column, which means that in crowded fields smear light curves tend to be significantly contaminated. 

The more precise method of halo photometry, which is the subject of this paper, uses the broad `halo' of scattered light around a saturated star to recover relative photometry, by constructing a light curve as a linear combination of individual pixel time series and minimizing a Total Variation objective function (TV-min). It has been employed for example on the Pleiades \citep{White2017} and the brightest-ever star on \kepler silicon, Aldebaran \citep[$\alpha$\,Tau;][]{Farr2018}, recovering photometry with a precision close to that normally obtained from \ktwo observations of unsaturated stars. Unlike smear, this requires downloading data out to a 12--20~pixel radius around each star, and has accordingly only been possible for stars that were specifically proposed and targeted with apertures optimized for this method, plus a small number of other stars for which this is fortuitously the case. The pixel requirements for this are sufficiently low that, with the help of the \ktwo Guest Observer office, such apertures were obtained for most of the bright targets from Campaign~4 onwards.

In this Paper we describe numerical experiments testing the TV-min method and extending it to generalizations with different exponents and timescales. We show that the method as previously employed applying standard TV-min is suboptimal, and gain a modest improvement from taking finite differences close to the timescale of \ktwo thruster firings. We also document the main changes in the halo data reduction pipeline, \texttt{halophot}, with respect to previous releases. 
We go on to present a complete catalog of long-cadence \ktwo halo light curves, which we have made publicly available. We have employed halo photometry on all stars targeted with appropriate apertures, and have done a preliminary characterization of interesting astrophysical variability. These include oscillating red giants, pulsating and quiet main-sequence stars, and eclipsing binaries, many of which are among the brightest objects of their type to have been observed with high-cadence space photometry. We are convinced that this diverse catalog of high-precision light curves will be useful for a range of astrophysical investigations.

\section{Halo Photometry Method}
\label{method}

The `TV-min' halo method was first described by \citet{White2017} and applied to the Pleiades' Seven Sisters. It was also applied to Aldebaran with further developments by \citet{Farr2018}. In this Section we will discuss some improvements made to the halo method since those publications, and describe tests of the method using saturated and unsaturated targets. 

We follow the Optimized-Weight Linear (OWL) photometry concept described by Hogg \&  Foreman-Mackey (2014, unpublished: preprint  \href{https://github.com/davidwhogg/OWL/}{\nolinkurl{github.com/davidwhogg/OWL/}}) in our assumptions. We assume that a star has a wide PSF sampled by many pixels with different sensitivities. This PSF varies at most to a small extent in time. The star moves around on the detector within a small region. We assume that our time series consists of many epochs sampled with a nearly even cadence. We do not wish to rely on metadata describing the spacecraft motion, pixel gains, PSF variations or other noise processes, at least at this stage.

Because photometry is a linear operation, any estimator of the flux is necessarily a weighted sum of pixel values. We choose these weights to be time-invariant but note that this strong constraint is not necessary in general. Allowing these weights to vary in time is a possible extension of this method to non-stationary noise processes, but we do not explore this further in this work. In OWL and here, we search for a linear combination of pixels that is invariant with respect to the the noise processes but accurately preserved astrophysical signals.

The additional constraint beyond the OWL axioms is that some pixels are saturated, so that Simple Aperture Photometry (SAP) is inadvisable. Instead the measurements are made using the unsaturated pixels at the wings of the broad and structured PSF, with counts $p_{ij}$ where pixels are indexed by $j$ and epoch by $i$. We construct a light curve as a linear combination of these time series with weights $w_j$, so that flux $f_i$ at epoch $i$ is

\begin{equation}
    f_i \equiv \sum_j {w_j p_{ij}}.
\end{equation}

\noindent In our updated pipeline presented here, the weights are chosen to minimize an objective function

\begin{equation}
    Q_{k,\delta} \equiv {\sum_{i>\delta}{|f_i - f_{i-\delta}|^k}},
\end{equation}

\noindent with an integer lag parameter $\delta$ and an integer $L_k$ norm, subject to the constraints 

\begin{align}
\forall_j w_j &> 0\\
\sum_{i=1}^{N} f_i  &= N.
\end{align}

\begin{figure*}
\plotone{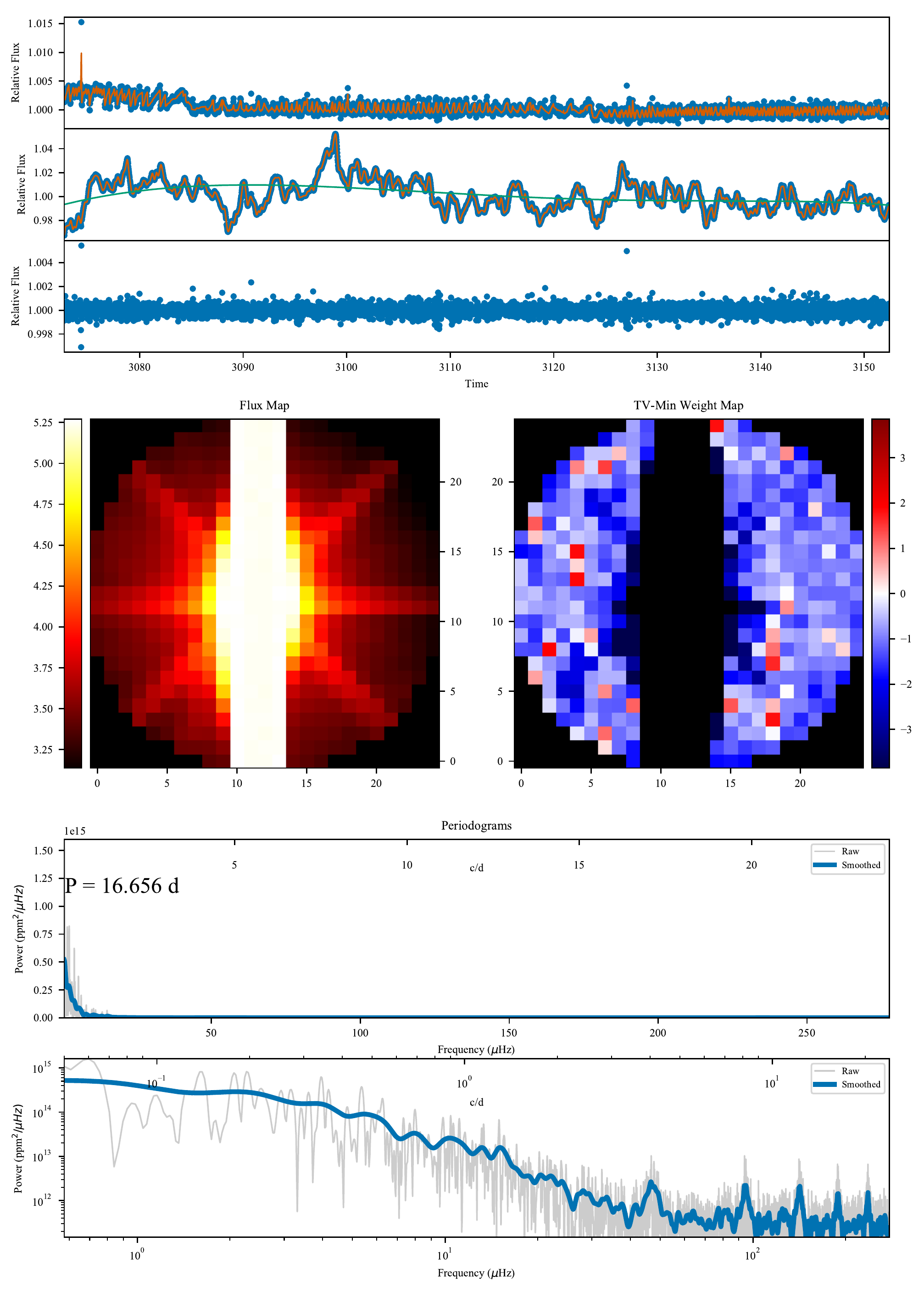}
\caption{Summary plots for \textsc{k2sc}-corrected final halo light curve for $\rho$~Leonis. The top three panels illustrate \textsc{k2sc} systematics correction: at the top, flux minus the GP time trend (blue dots) with GP $x,y$ trend superimposed (orange line); in the middle, flux minus GP $x,y$ components with GP time trend superimposed, and in green, a fifteenth-order polynomial trend; at the bottom the `whitened' light curve with flux minus both GP components. Middle two panels: log-flux map (left) and halo log-weight map (right). Bottom two panels: Lomb-Scargle power spectra \citep{lomb,scargle} in linear (top) and log (bottom) scales of the residuals of the corrected light curve minus the long term polynomial trend. Plots of this form are available on MAST for all long-cadence stars (\dataset[10.17909/t9-6wj4-eb32]{\doi{10.17909/t9-6wj4-eb32}}), together with similar plots for all short-cadence stars but without \textsc{k2sc}. The period at maximum power (16\,d) is marked on all plots of this form; in $\rho$~Leo, variability is attributed to red noise and a 26.8~d rotation period \citep{Aerts2018,bowman19}.}
\label{fig:rholeo}
\end{figure*}

This is a classic convex optimization program with constraints, which we solve with the \texttt{scipy} \citep{scipy} L-BFGS-B nonlinear optimization code \citep{lbfgsb}. $Q_{k,\delta}$ has analytic derivatives with respect to $w_j$ \citep[calculated with \texttt{autograd};][]{autograd}, and it is therefore extremely fast to optimize and converges well on a global solution. In practice, for computational reasons we optimize over parameters $\tilde{w}_j$ such that $w_j = \text{softmax}(\tilde{w}_j) = \exp{\tilde{w}_j}/\sum_j(\exp{\tilde{w}_j})$, where $\text{softmax}$ is the normalized exponential function. This satisfies the constraint that $\forall_j w_j > 0$, and while this also constrains their sum to be unity, we renormalize $f$ to satisfy its normalization constraint before calculating the objective function and this additional constraint is removed again. Weight maps displayed in Figures~\ref{fig:rholeo},~\ref{fig:98tau} and~\ref{fig:etacnc} display $w_j$ and not $\tilde{w}_j$. 

The objective function $Q_{k,\delta}$ is the $L_k$ norm on a `lagged' finite difference with a lag parameter $\delta$. For $k = 1$ and $\delta = 1$, $Q_{1,1}$ is the standard Total Variation objective (TV) used in previous halo papers \citep[e.g.][]{White2017,Farr2018}, and can be seen as the L1 norm on the derivative of $f$ or as a discrete approximation to its arc length. The L2 Variation (L2V) with $k=2$ is sometimes referred to in image processing literature as the `smoothness' regularizer, as it seeks to penalize large gradients without necessarily making them sparse. While $k$ does not have to be an integer in principle, in this implementation we have chosen to restrict our analysis to $k \in \{1,2,3\}$. The lag parameter $\delta$ allows for flexibility in modelling systematics occurring at different timescales from epoch-to-epoch, and we investigate its effects below. The order parameter $k$ allows for flexibility in how sensitive we are to normally-distributed versus long-tailed noise. For convenience in the rest of this paper, we will refer to the $k=1$ case as TV, the $k=2$ case as L2V, and the $k=3$ case as L3V. As the sampling in K2 is close to uniform but not perfectly uniform, some finite differences actually skip two or three cadences, but these are a small contribution to the final objective function; for very irregularly sampled data, it may be valuable to interpolate onto a uniform grid.

\citet{parker} in their work on the saturated \ktwo observations of Titan optimized an objective function equivalent to $Q_{2,1}$ with a \emph{second}-order finite difference $2f_i -f_{i-1} -f_{i+1}$, noting that first-order differences are sensitive to linear trends while second-order differences are invariant. We nevertheless choose to use a first-order finite difference, on the grounds that long-term astrophysical trends on the timescale of a \ktwo Campaign cannot be straightforwardly distinguished from systematics, and that the short-timescale noise performance of optimizing $Q_{2,1}$ with respect to first-order differences was superior in our numerical experiments. 

Unlike other methods for calibrating \kepler systematics, other than the value of~$\delta$, no knowledge of the spacecraft motion or the behaviour of an ensemble of other stars is used to inform our algorithm. The signal and the noise are jointly estimated from the data. The method is both self-calibrating, and is independent of the details of the systematics it is calibrating, operating on the assumption that a single signal is present across many individual time series which otherwise are contaminated by noise. 

It is therefore likely that significant improvements can be made to the method by including cotrending basis vectors with mean zero and whose weights are allowed to be negative, which would represent systematics which are common to all pixels in the halo aperture and therefore masquerade as signal. Any linear combination of convex objective functions is itself convex, and future extensions to the method could apply combinations of different lags and orders to better represent systematics occurring on different timescales (e.g. thruster firings, red noise) and with different levels of smoothness.


In addition to expanding the range of possible objective functions, we have also added a feature `\texttt{deathstar}' to deal with contamination. Clusters of pixels are identified with the \textsc{dbscan} algorithm \citep[Density-Based Spatial Clustering of Applications with Noise;][]{dbscan}, and we join these clusters with the watershed-based image segmentation algorithm from \textsc{k2p2} \citep{k2p2}. Clusters other than the target star identified by this algorithm are identified as possible background sources and removed from the target pixel file before processing. Other than this, we have adopted less-aggressive quality flagging, having found that many epochs were being classified as bad quality for spurious `cosmic ray' events, which were actually caused by a combination of saturation and spacecraft motion. We instead chose to iteratively sigma-clip outliers and use the \texttt{lightkurve} \citep{lightkurve} default quality mask.

While the halo procedure produced a fairly clean light curve in most cases, there were nevertheless residual systematic errors related to spacecraft motion. In order to correct these, we employed the \textsc{k2sc} code \citep{Aigrain2015,k2sc}, which simultaneously models a light curve as a 3D Gaussian Process (GP) in time and predicted position (the K2 standard data product \textsc{pos\_corr}) in pixels $(x,y)$. The model prediction in time for fixed position is then a nonparametric model of the stellar variability, and the prediction for the $x,y$ component evaluated for fixed time represents the pointing systematics. We subtracted the systematics model from the input fluxes to obtain a final corrected flux, which is the time series we use and recommend for science. Campaigns~9, 10, and~11 were observed in two blocks each, denoted C91/C92, C101/C102 and C111/C112 by the \ktwo Team. The target pixel files for C91, C92, and C101 include no position information. As a result \textsc{k2sc}-corrected data are not available for these targets.

\subsection{Choosing the Objective Function}
\label{sec:objective}

In order to choose the values for $k$ and $\delta$ in our objective function, we used the system 36~Ophiuchi (Guniibuu, $V= 5.08$), a K1/K2/K5 active main sequence triple system consisting of the lowest-mass main sequence stars in the sample of stars with halo apertures. Very little high frequency variability is detected or predicted. It was also observed at short cadence. We chose the 6.5~hour Combined Differential Photometric Precision \citep[CDPP,][]{cdpp} as implemented in \texttt{lightkurve} as a proxy for the `noise' in a light curve, with lower being better. 

\begin{figure*}
\plotone{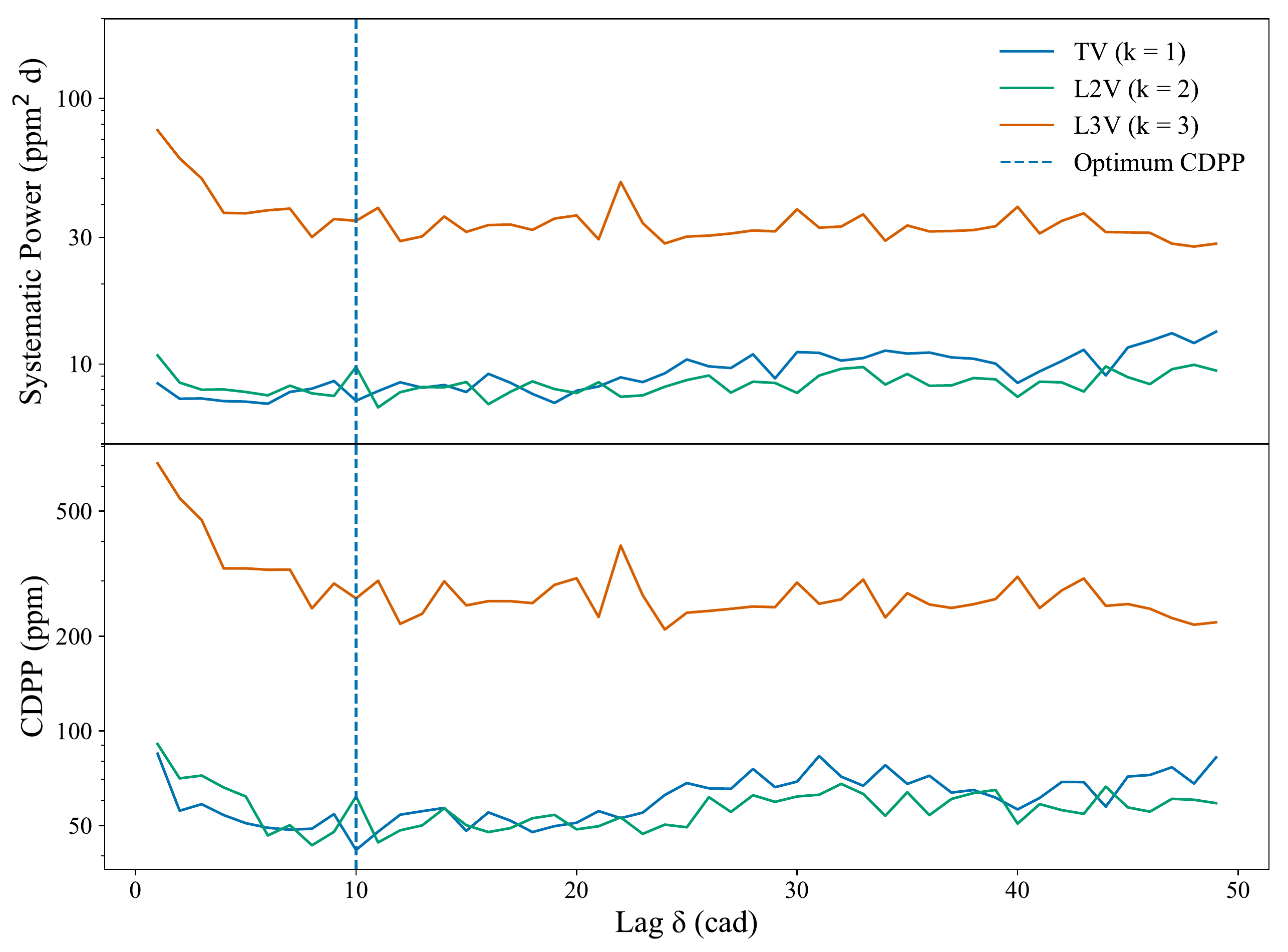}
\caption{Behaviour of long-cadence 6.5~hour CDPP (bottom) and $4 c/d$ systematics power (top) for the quiet dwarf 36~Ophiuchi as a function of lag parameter $\delta$. CDPP shows a minimum for L1 norm and $\delta=10$, i.e. for objective function $Q_{1,10}$, which is marked with a blue dashed vertical line. This does not correspond to an optimum in systematic power, which is slightly lower for smaller $\delta$. Nevertheless, we have chosen $\delta=10$ for the light curves in this catalog because of its improvement in overall CDPP as a measure of planet detection efficiency and overall light curve quality.}
\label{cdpps_lc}
\end{figure*}

\begin{figure}
\plotone{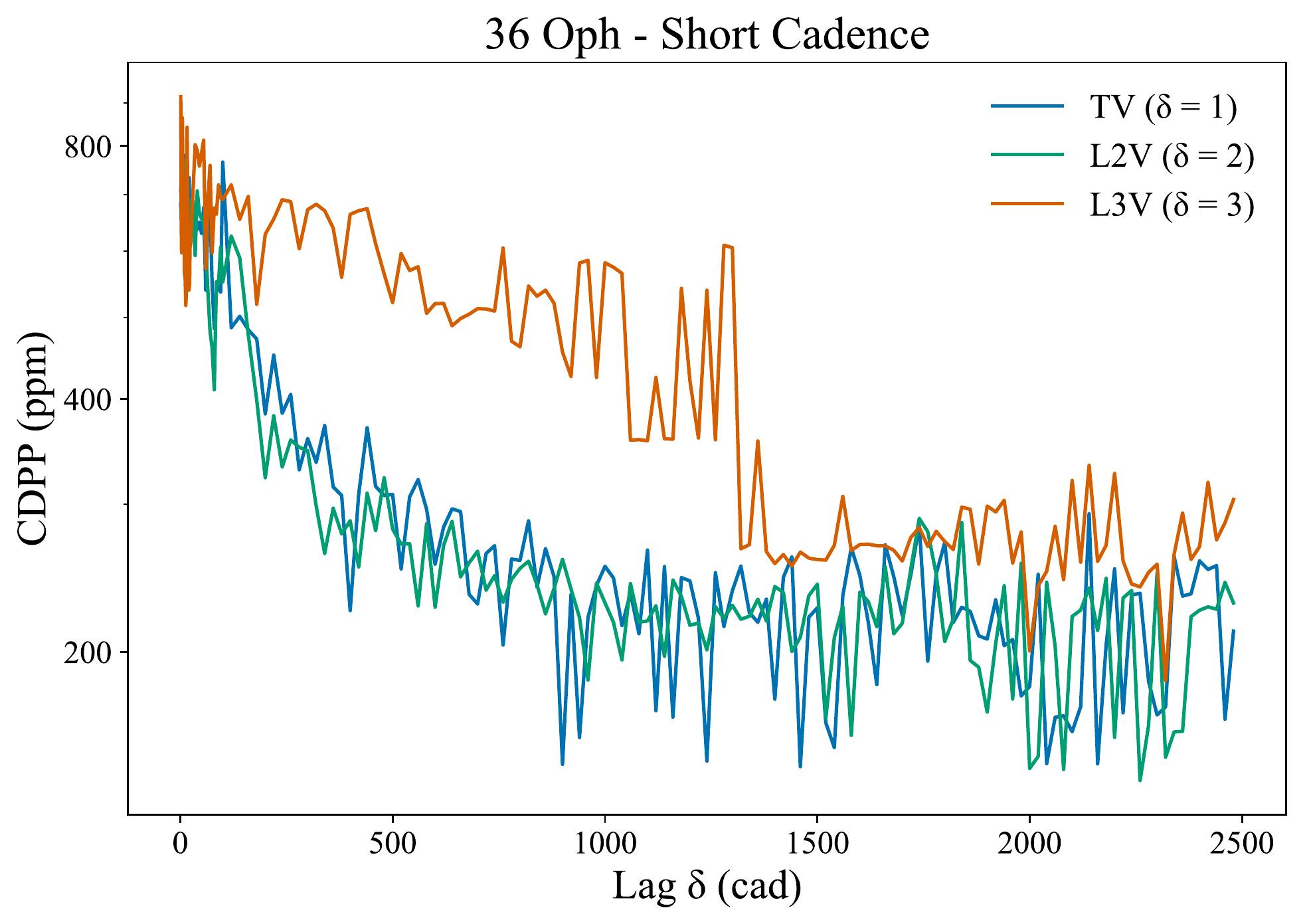}
\caption{Behaviour of short cadence 6.5~hour CDPP for the quiet dwarf 36~Ophiuchi as a function of lag parameter $\delta$. CDPP continuously improves for higher lags and shows no strong differences between L1 and L2~norms, while L3 performs poorly.}
\label{cdpps_sc}
\end{figure}

We calculated halo light curves of 36\,Oph and their CDPPs for $k \in \{1,2,3\}$, and $\delta \in [1,50]$ for long cadence and for various values of $\delta \in [1,2500]$ for short-cadence data. The results are displayed in Figures~\ref{cdpps_lc} and~\ref{cdpps_sc}. We found that for long-cadence data, $k=1$ (TV) and a lag $\delta=10$ provide the best CDPP, though not dramatically better than a range of values from $\sim 8-20$. As this is around the 12~cadence thruster firing period, we can understand the optimum as suppressing systematics on the same timescale as they occur. On the other hand, for short-cadence data, performance at short lags is very poor but the method performs similarly for $k \in \{1,2\}$ with slow improvement with larger $\delta$, and performs very poorly for $k=3$ at all lags. 

We accordingly use a lag $\delta=10$ for all long-cadence light curves, and a lag $\delta=300$ for short cadence for consistency in timescale with the long-cadence processing.



\subsection{Benchmarking}
\label{sec:benchmarking}

As the halo method is the only available means of obtaining light curves of stars as bright as in our sample, and they are ubiquitously found to be variable, it is difficult based on this sample alone to determine the accuracy and precision of the light curves obtained. While \citet{Kallinger2018} have found agreement between the \citet{White2017} halo observations of Atlas and their BRITE-Constellation observations, the BRITE observations have a lower precision and cannot be obtained for most of the stars in our sample. 

\begin{figure*}
\plotone{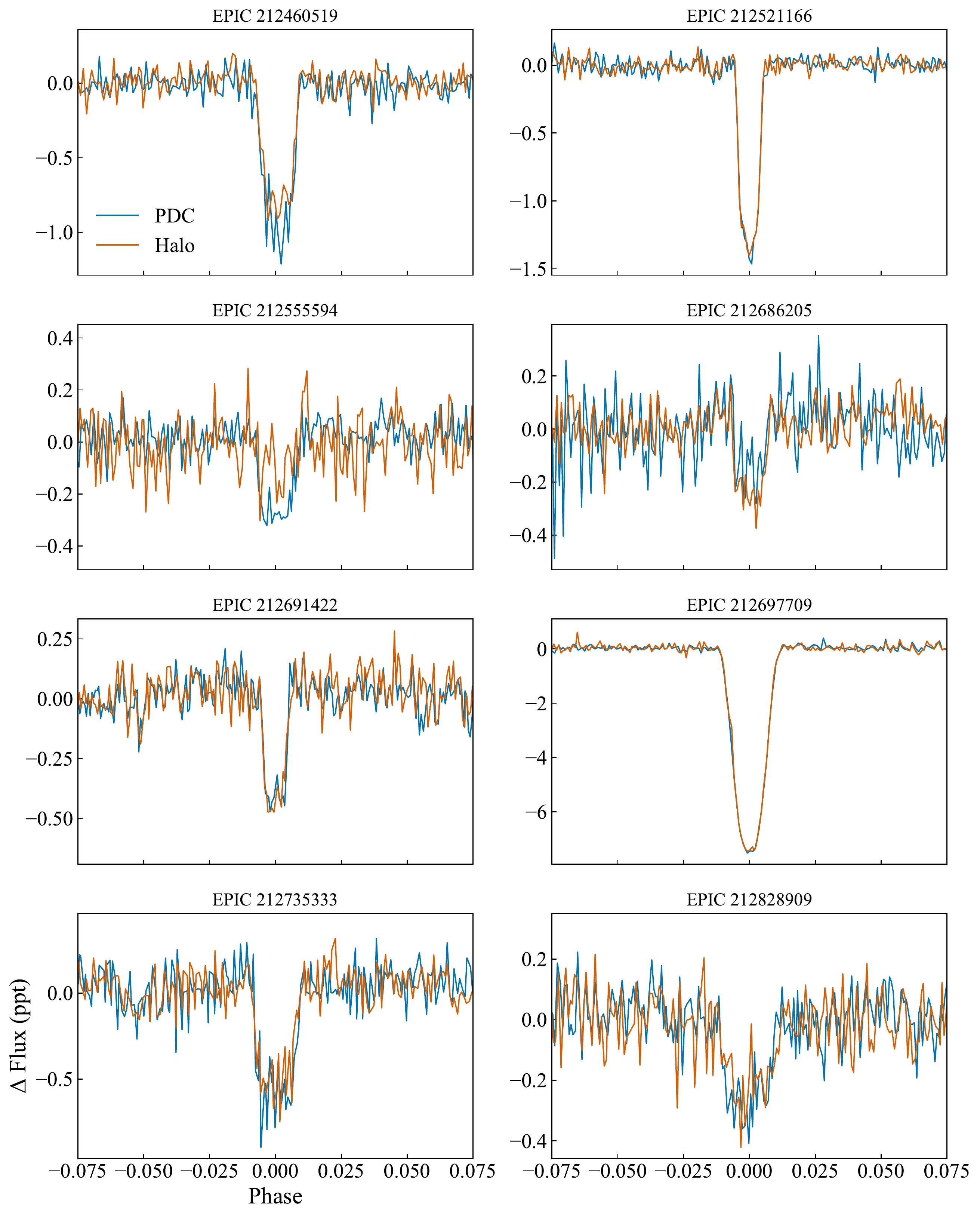}
\caption{The eight transiting single-planet systems in \ktwo Campaign~6 in the magnitude range $11.5 < Kp < 12.5$, with PDC light curves (blue) and TV-min light curves (orange) overlaid. These have been identically \textsc{k2sc}-corrected, whitened, outlier-clipped, folded and binned as described in Section~\ref{sec:benchmarking}. The depths and shapes of the transits agree closely except for EPIC~212460519, for which the TV-min transit is slightly shallower, and EPIC~212555594, for which TV-min is significantly shallower.}
\label{fig:planets}
\end{figure*}

We want to compare the photometric precision obtained to that from SAP and normal calibration pipelines, and ascertain whether we systematically distort the scale of variation or the power spectrum of variability. In order to do this, we take the sample of stars with $11.5 <  Kp < 12.5$ from \ktwo Campaign~6, for which \textsc{k2sc} light curves are available, choosing~2466 stars that are as bright as possible without saturation. The planets in this campaign are well characterized \citep[e.g.][]{Pope2016planets}, and eight singly-transiting systems are known in this magnitude range. We take the entire target pixel file without using any aperture restriction, and run TV-min with $\delta = 10$ for each of these planets and compare these to light curves from the PDC pipeline. In both cases, we correct residual systematics with \textsc{k2sc}, prewhiten with the GP time trend model, clip $3\,\sigma$ upwards outliers, and normalize the final fluxes to unity. These are then folded on the known transit period and zero epoch as tabulated in the NASA Exoplanet Archive \citep{2013PASP..125..989A}, and the folded light curves are binned in 3-epoch bins to reduce white noise in the comparison. The results are displayed in Figure~\ref{fig:planets}.

\begin{figure}
\plotone{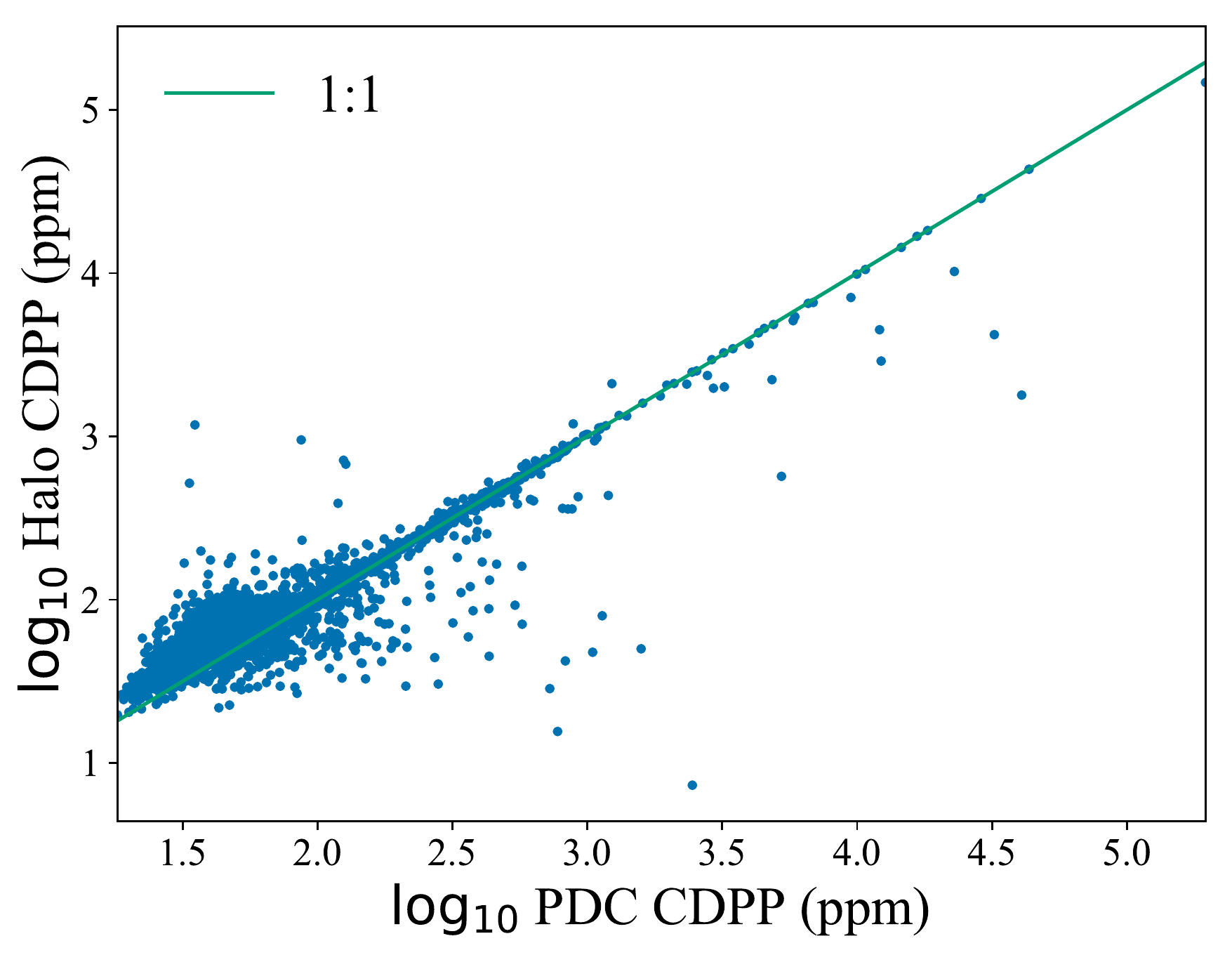}
\caption{Correlation diagram of the \texttt{lightkurve}-computed 6.5\,hr CDPP for \ktwo Campaign~6 stars in the magnitude range $11.5 < Kp < 12.5$, as processed with the PDC pipeline ($x$-axis) and TV-min pipeline ($y$-axis), both after correction and whitening with \textsc{k2sc}. The severe outliers where halo significantly outperforms PDC are shown by individual inspection to consist of stars for which there is contamination, or for which the SAP aperture assigned by the \kepler pipeline is significantly smaller than the PSF.}
\label{fig:halovspdc}
\end{figure}

We now seek to establish the global noise properties of the whole unsaturated sample, and compare these to PDC. We process all~2466 stars with TV-min and $\delta = 10$, using all pixels in the TPF unmasked. Because these stars are so bright and the TPFs so small, in the great majority  of cases we do not expect significant contamination, and this is a way of testing how well the weights assigned by TV-min match the flux distribution over pixels. For each light curve we calculated the 6.5\,hr CDPP proxy with \texttt{lightkurve} as a measure of SNR, and we plot the results of the two pipelines against one another in Figure~\ref{fig:halovspdc}. We see that a significant number of stars have high PDC CDPP but low TV-min CDPP, which raises the possibility that these are variables for which halo is overcorrecting. We found by inspection of the weightmaps and \kepler pipeline aperture masks that these mostly consist of stars for which the SAP aperture is significantly smaller than the PSF. In this case, by ignoring the pipeline apertures, \texttt{halophot} is in fact generating significantly better light curves. Over all stars, we found that the fractional enclosed halo weight in the \kepler pipeline aperture is only $0.19 \pm 0.11$, which suggests that in fact the pipeline apertures are systematically smaller than optimal for stars of this magnitude, and that TV-min is using information in the fainter pixels to help correct systematics.

\begin{figure}
\plotone{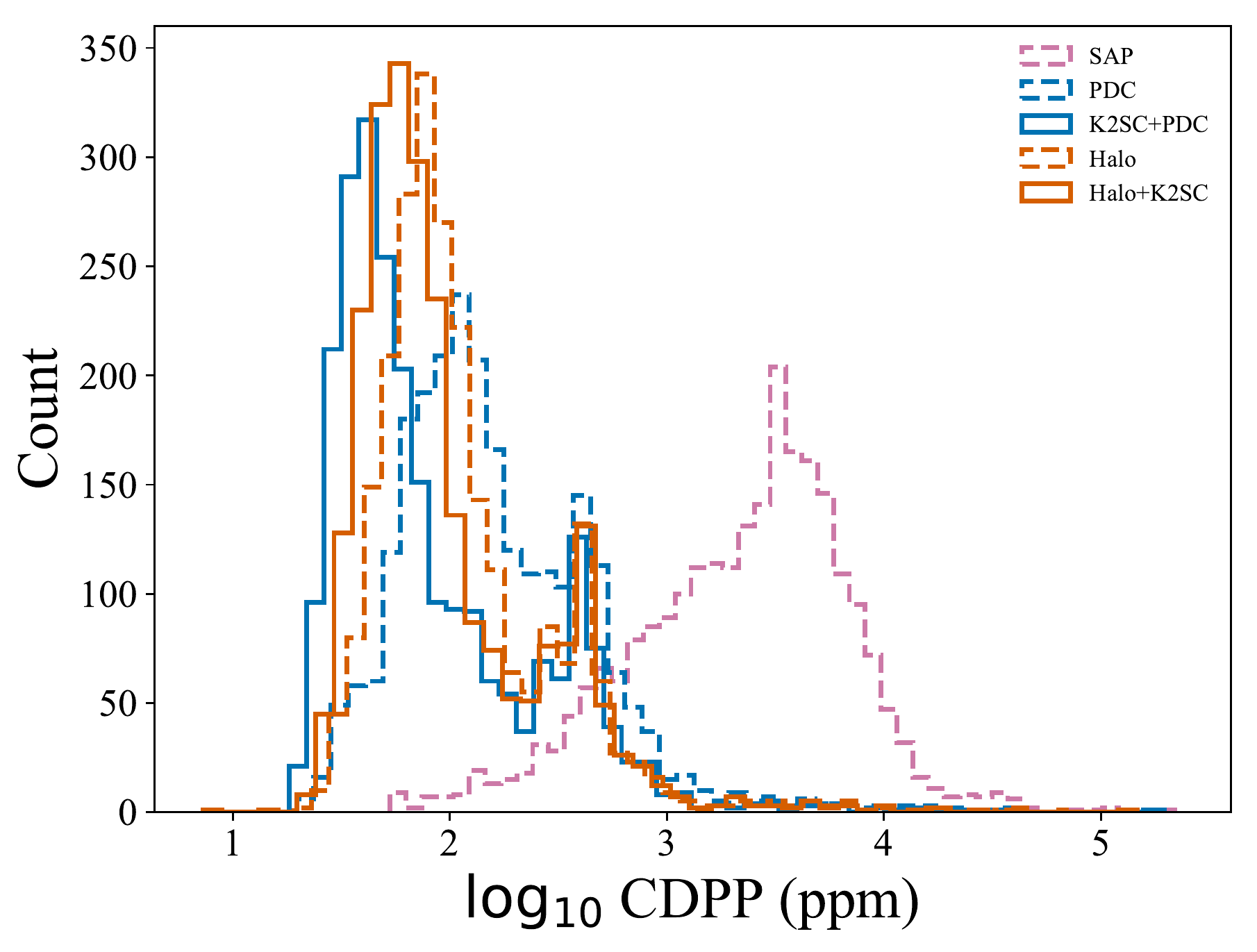}
\caption{Histograms of the \texttt{lightkurve}-computed 6.5\,hr CDPP for five different pipelines applied to \ktwo Campaign~6 stars in the magnitude range $11.5 < Kp < 12.5$: SAP (purple dashed), PDC with (blue solid) and without (blue dashed) \textsc{k2sc}, and TV-min with (orange solid) and without (orange dashed) \textsc{k2sc}.}
\label{fig:cdpphists}
\end{figure}

Histograms of the CDPPs of the SAP, PDC and halo light curves with and without \textsc{k2sc} are displayed in Figure~\ref{fig:cdpphists}. We see that both halo and PDC significantly outperform SAP, with halo performing better than PDC with no additional correction. Nevertheless, after \textsc{k2sc}, we found that the best PDC light curves have a smaller CDPP than the best similarly pointing-corrected halo. We conjecture that PDC with its improved calibration for common-mode systematics and blended/background light is correcting for effects that halo, as a single-star and instrument-agnostic method, does not.

\section{Sample}
\label{sec:sample}

\begin{figure}
\plotone{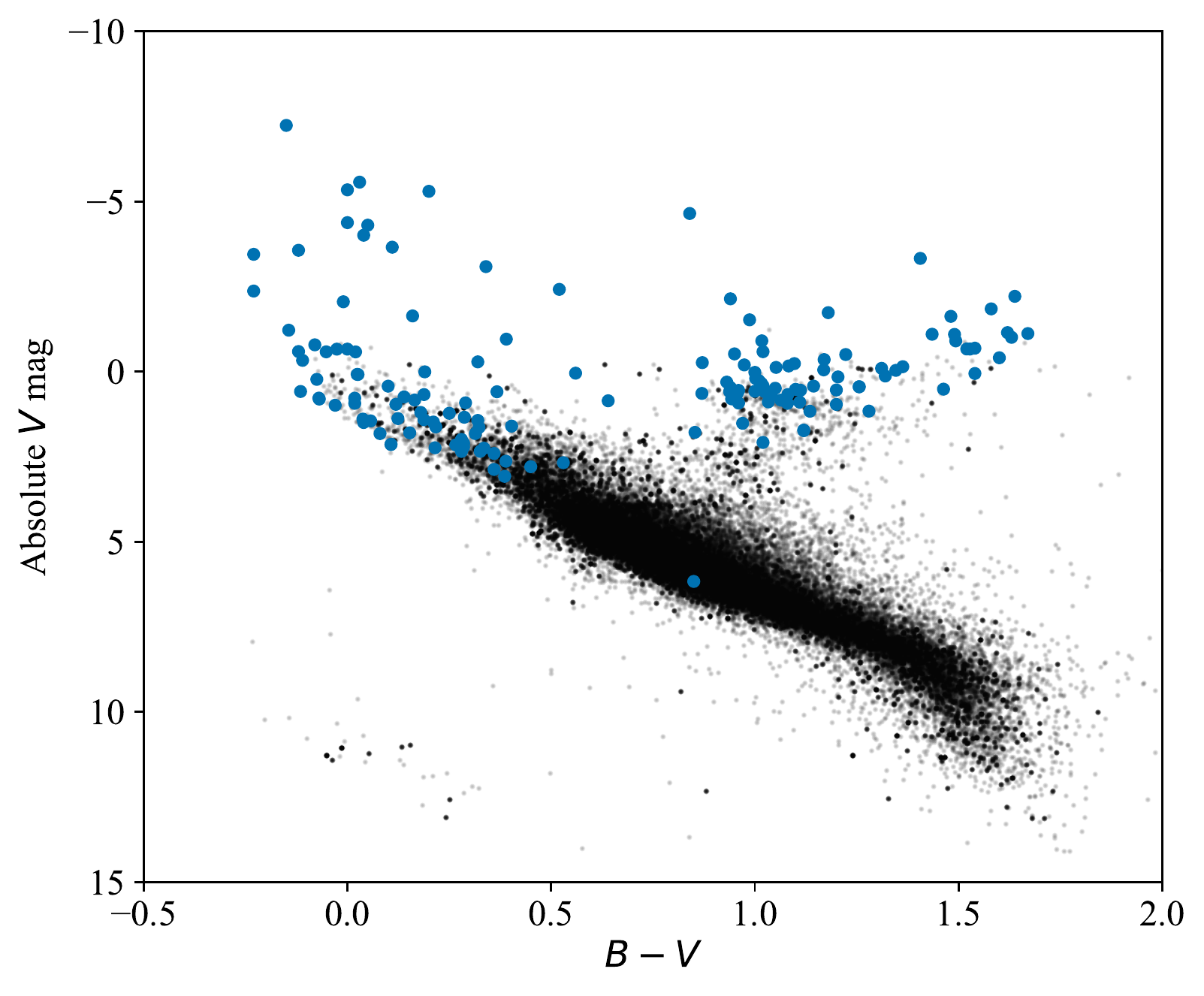}
\caption{$B$, $V$ color-magnitude diagram of the halo sample overlaid on a random subset of K2 stars with high-SNR \emph{Gaia} cross-matches, from the \href{https://gaia-kepler.fun}{\nolinkurl{gaia-kepler.fun}} sample, with $B$ and $V$ magnitudes drawn from the EPIC catalog \citep{huber16}. We omit the very-red carbon star HR~3541, whose $B-V$ color is 3.23. The halo sample is seen to be more intrinsically luminous than K2 stars overall, and includes the most intrinsically luminous star observed by K2, $\rho$~Leonis. An interactive version of this plot is available at \href{https://benjaminpope.github.io/data/cmd_halo.html}{\nolinkurl{benjaminpope.github.io/data/cmd_halo.html}.}}
\label{cmd_halo}
\end{figure}

The full sample of the 161 stars for which halo apertures were obtained is listed in Table~\ref{table_all}. A $B$, $V$ color-magnitude diagram is displayed in Figure~\ref{cmd_halo}, omitting the very-red carbon star HR~3541, whose $B-V$ color is 3.23. Following the successful pilot observations of the Pleiades B~stars in Campaign~4, we proposed halo photometry through dedicated K2 Guest Observer Programs from Campaign 6 onwards. Target selection was performed by cross-matching Hipparcos \citep{leeuwen07} with the K2 Ecliptic Plane Input Catalog \citep[EPIC,][]{huber16} and selecting all targets on silicon brighter than $Kp < 6$ on silicon. M~giants which pulsate with periods that are long compared to a K2 campaign were removed. We requested short-cadence observations for a small number of unevolved stars for which the expected timescales of oscillations cannot be sufficiently sampled with long-cadence data, such as for $\delta$~Sct stars whose maximum frequencies can exceed the long-cadence Nyquist limit.

 Some very bright stars were observed with conventional apertures as part of these programs, but we exclude them from the present discussion and data release, which is oriented towards targets only observable with halo photometry. 
We include $\alpha$\,Vir (Spica) and 69~Vir, which were observed in Campaign~6 without a halo aperture (in Campaign~17 Spica was re-observed, with a halo aperture). In Campaign~6 they were assigned normal apertures due an erroneous estimate of their \kepler magnitudes and simple aperture photometry performed extremely poorly, so we have processed these data with the halo pipeline. The stars in Campaign~18 in our sample were also on-silicon in Campaign~5, but were not assigned apertures suitable for halo photometry in C5. A possible further extension of the present work would be to recover C5 light curves for these objects using smear and/or modified halo photometry.

Seven stars in Campaign~13 and one in Campaign~16 were assigned short-cadence halo apertures. For these targets we have provided both long- and short-cadence reductions. Following the analysis in Section~\ref{method} showing the insensitivity of short-cadence CDPP to lags longer than $\sim 100$\,cad and to $k \in {1,2}$, and for consistency with long cadence, we have adopted a 300~epoch lag (i.e. $30 \times$ the long-cadence lag of 10) and the L1 TV objective function. With their many time samples, the short-cadence stars are computationally intractable for the Gaussian Process model in \textsc{k2sc} and we present otherwise uncalibrated halo light curves.



Analyses for some of our sample have been previously published, we include their light curves in this data release: the Pleiades' Seven Sisters \citep{White2017}, $\alpha$~Tau \citep[Aldebaran;][]{Farr2018}, $\iota$~Lib \citep{Buysschaert2018}, and $\epsilon$~Tau \citep[Ain;][]{Arentoft2019}, as well as $\rho$~Leo, which was studied with halo pixels but without our objective functions \citep{Aerts2018}.

\section{Discussion}
\label{sec:discussion}

\subsection{Comparison with `Raw' Halo}
\label{raw}

\begin{figure}
\plotone{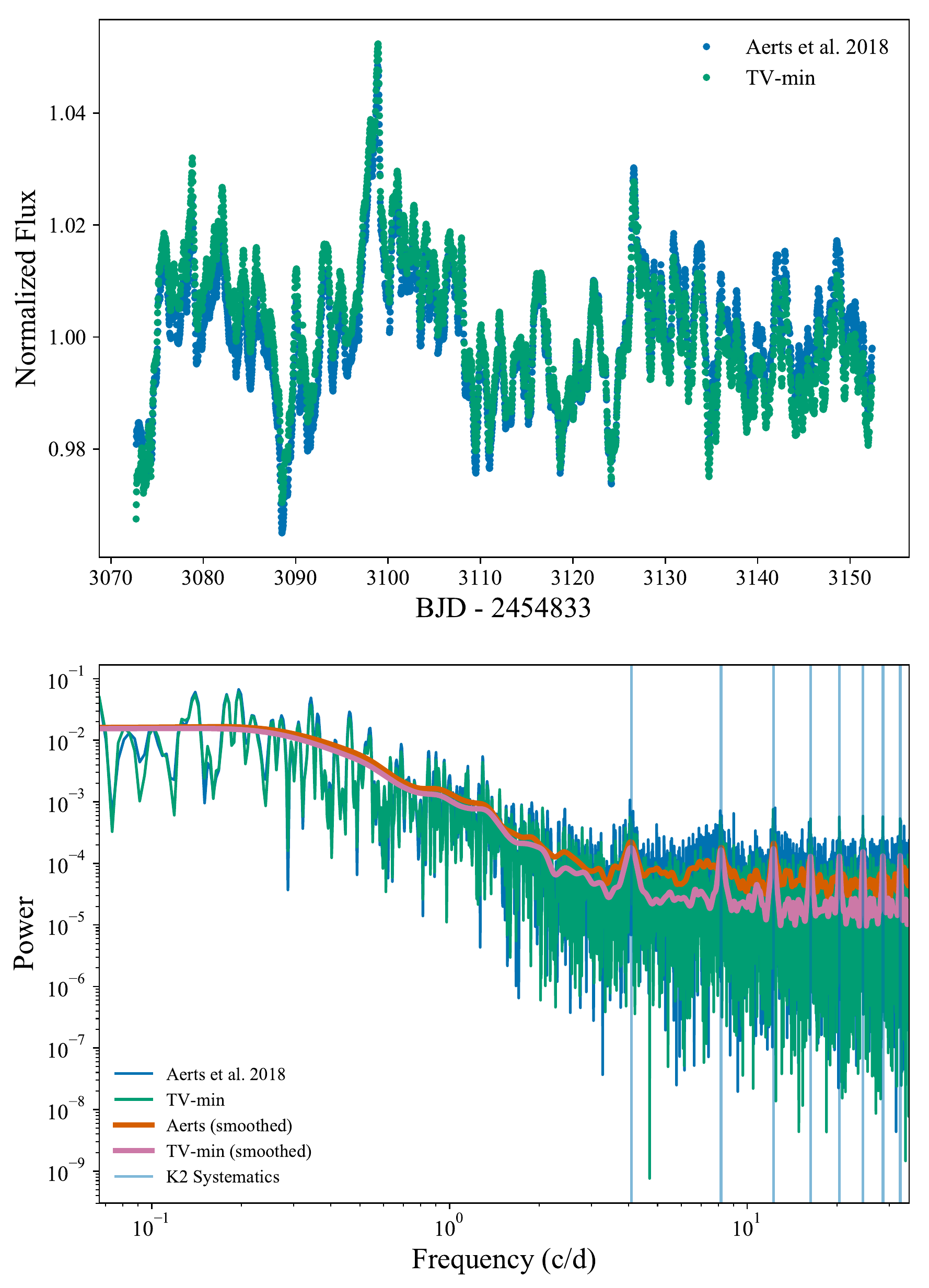}
\caption{Top: halo light curves of $\rho$~Leonis from \citet{Aerts2018} (blue) and TV-min from the present paper (green). Bottom: Lomb-Scargle power spectral densities of the Aerts (blue) and TV-min (green) observations, with smoothed power spectral densities overplotted in orange and purple respectively and the K2 thruster firing frequencies highlighted with pale blue vertical lines. There is excellent agreement between the light curves and power spectra at high frequencies, with some residual thruster firing systematics in both light curves, and a factor of $\sim 3$ lower white noise floor in the TV-min power spectrum.}
\label{rholeo_comparison}
\end{figure}

The blue supergiant $\rho$~Leonis, observed in Campaign~14, was studied with halo photometry but without the TV-min method by \citet{Aerts2018}. In that reduction, \citet{Aerts2018} used four different aperture masks to extract raw light curves, and detrended these for K2 systematics with \textsc{k2sc} and a polynomial to account for long-term drift. They detected photometric variability at the star's rotation period of 26.8~d and also multiperiodic low-frequency variability ($<1.5 \text{d}^{-1}$). The \textsc{k2sc} systematics and variability models, residuals, halo apertures, and periodograms are shown in Figure~\ref{fig:rholeo}, and a comparison with the \citet{Aerts2018} lightcurve in Figure~\ref{rholeo_comparison}. There is excellent agreement between the light curves produced by both methods. It is easiest to compare the methods in the power-spectral domain, where we see a reduction of only a few percent in the amplitude of oscillations in the TV-min and the \citet{Aerts2018} lightcurve; at high frequencies, both methods show significant residual systematics at the K2~thruster-firing frequencies, but the TV-min lightcurve shows a lower white noise floor by a factor of $\sim 3$. 

\subsection{Oscillating Red Giants}
\label{sec:rgs}
\begin{figure*}
\plotone{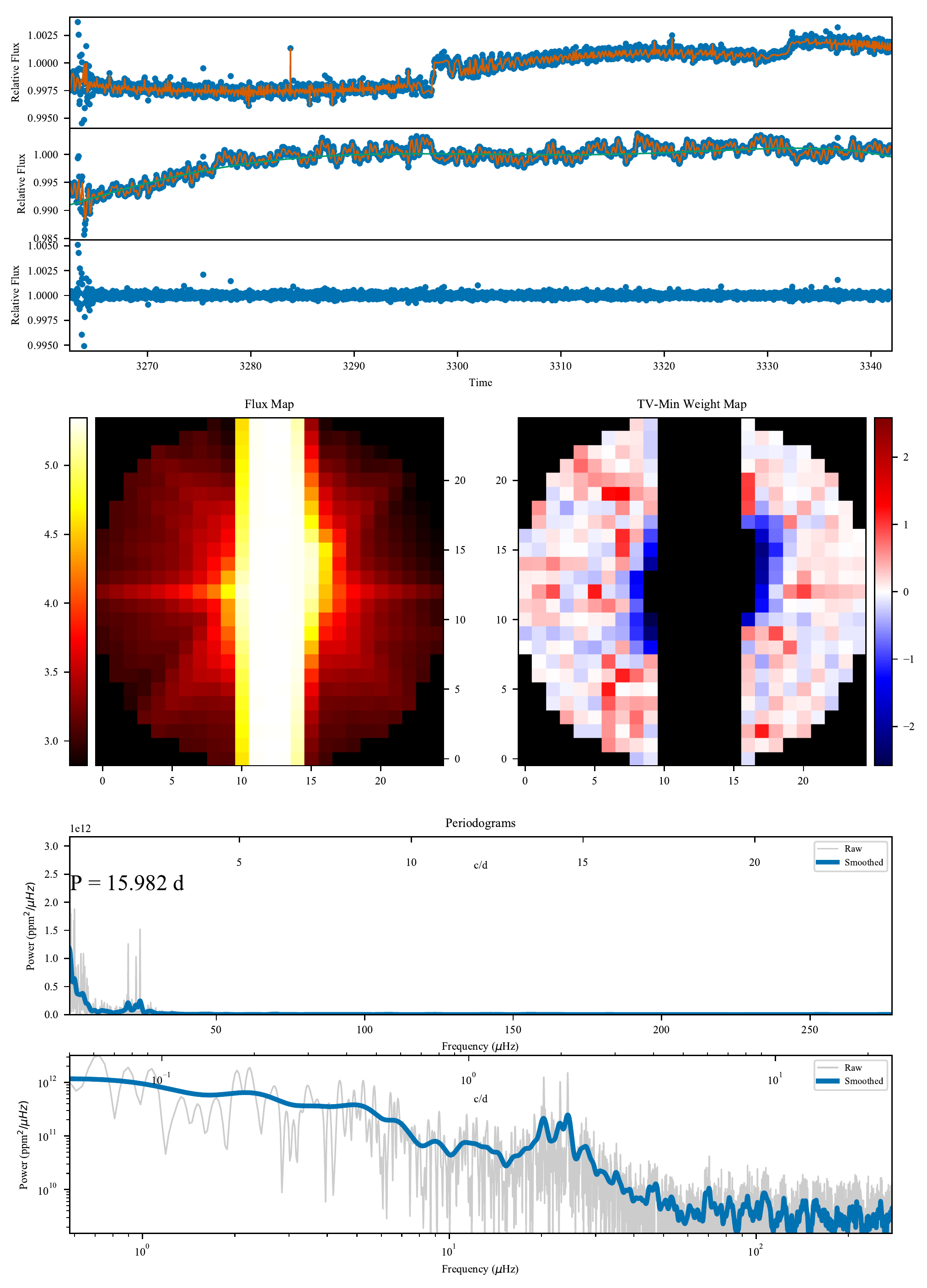}
\caption{Summary plots for \textsc{k2sc}-corrected final halo light curve for the red giant $\eta$~Cancri, in the same format as Figure~\ref{fig:rholeo}. Solar like oscillations are clearly detected with $\numax = 22.9 \pm 0.9$\,\muHz and $\Dnu = 2.7 \pm 0.03$\,\muHz.}
\label{fig:etacnc}
\end{figure*}

Thirty-one of the red giants in our sample have detectable stochastically-excited solar-like acoustic (p-mode) oscillations. In the asymptotic limit, these consist of a comb of modes separated by the large frequency separation \Dnu, approximately the sound-crossing-time of the star, with a Gaussian envelope centred on the frequency of maximum power \numax, which scales with the acoustic cutoff frequency at the star's surface. 
These \Dnu and \numax values can be used to constrain stellar fundamental parameters, such as radius, mass, and age \citep[e.g.][for a recent review]{2017A&ARv..25....1H}. Detailed studies of the deviations from the asymptotic limit for p-modes, e.g. due to acoustic `glitches', provide information on the He content and mixing processes at the bottom of the convective envelope \citep[e.g. ][]{Verma2019}. On the other hand, dipole mixed modes, which have a g-mode character in the inner regions of the star, fulfill an asymptotic period spacing determined by the buoyancy frequency inside the star. This spacing can be used to accurately determine the stellar evolutionary stage, and allows us to distinguish between hydrogen shell and core helium burning \citep{bedding2011}. 
Summary plots for a good example of such a star, $\eta$~Cancri, are shown in Figure~\ref{fig:etacnc}.

Using the Sydney pipeline \citep{Huber2009} with modifications to the extraction of \Dnu detailed in \citet{Yu2018}, we extract the global asteroseismic parameters \numax and \Dnu for the~31 red giants for which oscillations are detected with sufficient signal-to-noise. These parameters are listed in Table~\ref{rgs}; the stars are noted as showing `RG' variability in Table~\ref{table_all}, whereas this field is left blank for stars of luminosity class III for which oscillations are not unambiguously detected. High precision spectroscopy of these stars would permit detailed stellar modelling and the extraction of precise elemental abundances, which would make these stars useful as benchmarks for large spectroscopic surveys or testing detailed stellar models. This sample will be an addition to the~36 Gaia FGK benchmark stars \citep{gaiabenchmark1,gaiabenchmark3,2018RNAAS...2c.152J}, the 23 BRITE-Constellation asteroseismic red giants \citep{Kallinger2019}, and the~33 \kepler Smear Campaign spectroscopic benchmark red giants \citep{smearcampaign}. 

\subsection{Eclipsing Binaries}
\label{sec:ebs}

We have detected two eclipsing binaries in our sample: the previously-known EB HR~6773 and the new detection 98~Tau. After subtracting an EB model for HR~6773, we find additional variability consistent with SPB pulsations. 

\begin{figure*}
\plotone{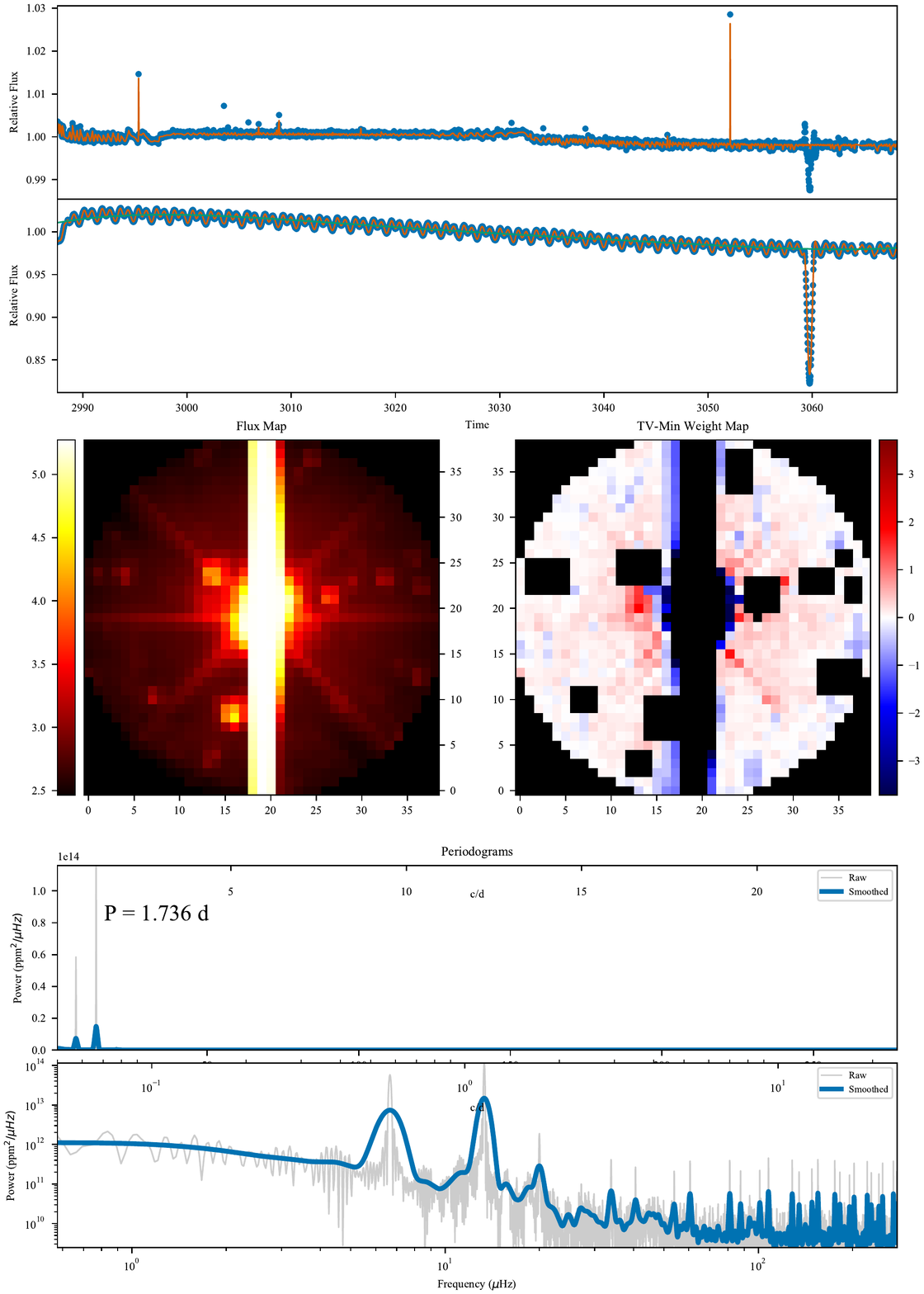}
\caption{Summary plots for \textsc{k2sc}-corrected final halo light curve for the eclipsing binary 98~Tauri, in a similar format to Figure~\ref{fig:rholeo}. Blacked-out pixels in the halo weights are background stars which were manually set to zero weight by hand. The residuals to the position and time GP are not shown, as the time GP fits poorly to the deep eclipse, though this did not adversely affect the pointing systematics model. The polynomial trend and Lomb-Scargle periodograms are conditioned on the out-of-transit points only.}
\label{fig:98tau}
\end{figure*}

The chemically-peculiar A0V star 98~Tau is of special interest for studies of surface inhomogeneity. We detected variability with a fundamental period of 1.74~d with twice as much power at the first harmonic ($P = 0.87$\,d), which is consistent with $\alpha^2$\,CVn chemical spot modulation from a rapidly-rotating star. This star also experiences a V-shaped transit of fractional depth 0.16, which for a 1.87\,\rsun typical A0V star implies a grazing eclipse by a stellar mass companion. There are an unusually high number of background stars in the same photometric aperture as 98~Tau, and these were not all detected by \texttt{deathstar} and significantly contaminated the resulting lightcurve. As a result it was necessary to manually flag these objects using the `interact' mode of \texttt{lightkurve}, as displayed in Figure~\ref{fig:98tau}. The eclipse is deep enough to be seen by eye in the diffuse light of 98~Tau using this interactive display, and is not associated with any of the background stars.

These systems contain variable stars in the brightest EBs in \ktwo, and are therefore unique targets for follow-up with smaller telescopes. With an eclipse to break degeneracies, models such as \texttt{starry} \citep{starry} has been shown to robustly and uniquely infer surface brightness maps from light curves. High-time-cadence photometry during transit, such as with CHEOPS \citep{cheops}, will reveal the spatial distribution of the star's chemical peculiarity or pulsation. 

\subsection{Other Variables}
\label{sec:variables}

Our dataset includes a rich variety of classical pulsators. We visually inspected the light curves and amplitude spectra to classify all non red-giant stars into traditional variability classes. We identify 23 stars showing $\delta$ Scuti pulsations and 20 with $\gamma$ Doradus pulsations, including 9~with hybrid $\delta$\,Sct/$\gamma$\,Dor variability; 14~slowly pulsating B~stars (SPB stars), 3~$\beta$ Cephei pulsators, and 3~Cepheids; as well as 3~O\,stars and 5~blue supergiants showing low-frequency variability \citep[as in][]{Aerts2018,bowman19}. In addition to this, the light curves of eight stars reveal rotational modulation, of which two have the characteristics of $\alpha^2$\,CVn chemical spot modulation. The classes we have determined for each star are listed in Table~\ref{table_all}. A detailed frequency analysis of the variability in each star will be presented in a forthcoming paper.

\section{Data Release and Open Science}

The software \texttt{halophot} that implements halo photometry as described in this paper is available under a GPLv3 license from \href{https://github.com/hvidy/halophot}{github.com/hvidy/halophot}. 

All light curves presented in this paper are available as High-Level Science Products from the Mikulski Archive for Space Telescopes (MAST)\footnote{\dataset[10.17909/t9-6wj4-eb32]{\doi{10.17909/t9-6wj4-eb32}}}.  They are also available, together with the source code that produced the survey sample and this manuscript, from \href{https://github.com/benjaminpope/k2halo}{github.com/benjaminpope/k2halo}.

\section{Conclusions}
\label{sec:conclusions}

We have presented an updated method for halo photometry, and used this to obtain light curves of 161~stars in \ktwo that were too saturated to be otherwise retrievable. These ubiquitously show variability, and we have presented global asteroseismic analysis of 31~red giants and variability classifications for all stars. This is a unique legacy sample for \ktwo, dramatically increasing the number of very bright stars that have been characterized with high-precision, rapid-time-cadence space photometry. We hope that our data release will be used for a variety of astrophysical investigations.

Some of the objects presented here are the subject of more detailed work in preparation, namely $\alpha$\,Vir (Spica), interferometry and asteroseismology of the Hyades giants, and main-sequence stars with self-driven nonradial modes.

The sample of K2 bright stars presented here only includes those with halo apertures. While some others are available conventionally, many were not assigned target pixels and were not downloaded at all. Smear photometry has been used to recover the brightest otherwise-unobserved stars in nominal \kepler \citep{smearcampaign}, and this can also be done in K2, although the sample is much smaller due to allocation of halo apertures and the systematics correction is more challenging. A natural extension of both pieces of work would be to produce smear light curves of all bright stars without halo apertures in K2, which would finally make the \kepler extended mission magnitude-complete at the bright end. 

The halo method naturally extends to other contexts where simple aperture photometry is not possible, such as for saturated stars observed by the Transiting Exoplanet Survey Satellite \citep[TESS;][]{tess}. Although the saturation limit is brighter ($T_{mag} \sim 6$) and this problem accordingly affects fewer stars and less badly, there are stars such as $\alpha$~Centauri and $\beta$~Hydri where the bleed column reaches the edge of the chip and a SAP light curve is irrecoverable. We expect that TV-min halo photometry will therefore be important in ensuring that TESS can observe the very brightest stars.

There are directions for improvement of the halo method itself, and for applying it beyond \kepler/\ktwo and TESS. It remains to be seen how well the method of optimizing convex objective functions can deal with significantly varying PSFs, such as from ground-based observations. The rapidly varying and moving seeing-limited PSF couples to flat field errors as is the case with \kepler, and leads to severe short-timescale instrumental noise. Self-calibration by the halo method, or a similar method, may permit improvements in ground-based photometry. Likewise, there may be other convex objective functions, including linear combinations of currently-used objective functions, which offer superior performance, for example by using combinations of different lagged functions to suppress systematics occurring at different timescales. The remaining unexplored space of convex objective functions may offer significant improvements on existing self-calibration techniques in high-cadence photometry and related problems in astronomy.

\section*{Acknowledgements} 

The halo apertures were kindly provided by the K2 team as part of the Guest Observer programs GO6081-7081, GO8025, GO9923, GO10025, GO11047-13047, GO14003-16003, and GO17051-19051, and as a Director's Discretionary Time program in Campaign 4 as GO4901. We gratefully acknowledge financial support by the National Aeronautics and Space Administration through K2 Guest Observer Programs NNX17AF76G, 80NSSC18K0362, and 80NSSC19K0108, which has been essential in bringing this project to fruition.

This work was performed in part under contract with the Jet Propulsion Laboratory (JPL) funded by NASA through the Sagan Fellowship Program executed by the NASA Exoplanet Science Institute. BJSP also acknowledges the financial support of the Clarendon Fund and Balliol College. TRW acknowledges the support of the Australian Research Council (grant DP150100250) and the Villum Foundation (research grant 10118). SA acknowledges support from the UK Science and Technology Facilities Council (STFC) under grants ST/N000919/1, ST/S000488/1, and ST/R004846/1. CA received funding from the European Research Council (ERC) under the European Union’s Horizon 2020 research and innovation programme (grant agreement N$^\circ$670519: MAMSIE) and from the KU\,Leuven Research Council (grant C16/18/005: PARADISE).

This project was developed in part at the Building Early Science with TESS meeting, which took place in March 2019 at the University of Chicago.

BJSP acknowledges being on the traditional territory of the Lenape Nations and recognizes that Manhattan continues to be the home to many Algonkian peoples. We give blessings and thanks to the Lenape people and Lenape Nations in recognition that we are carrying out this work on their indigenous homelands. We would like to acknowledge the Gadigal Clan of the Eora Nation as the traditional owners of the land on which the University of Sydney is built and on which some of this work was carried out, and pay their respects to their knowledge, and to their elders past, present, and emerging.

This research made use of NASA's Astrophysics Data System; the SIMBAD database, operated at CDS, Strasbourg, France. Some of the data presented in this paper were obtained from the Mikulski Archive for Space Telescopes (MAST). STScI is operated by the Association of Universities for Research in Astronomy, Inc., under NASA contract NAS5-26555. Support for MAST for non-HST data is provided by the NASA Office of Space Science via grant NNX13AC07G and by other grants and contracts. We acknowledge the support of the Group of Eight universities and the German Academic Exchange Service through the Go8 Australia-Germany Joint Research Co-operation Scheme. This work made use of the gaia-kepler.fun crossmatch database created by Megan Bedell.

\software{\texttt{halophot} \citep{White2017}; \textsc{k2sc} \citep{Aigrain2015,k2sc}; \texttt{lightkurve} \citep{lightkurve}; \texttt{autograd} \citep{autograd}; \textsc{dbscan} \citep{dbscan}; IPython \citep{PER-GRA:2007}; SciPy \citep{scipy}; and Astropy, a community-developed core Python package for Astronomy \citep{astropy}.}



\bibliography{ms}

\newpage
\appendix

\startlongtable
\begin{deluxetable}{ccllccl}
\tablecaption{All stars observed with halo photometry in K2.\label{table_all}}
\tablehead{\colhead{Name} & \colhead{EPIC} & \colhead{Spectral} & \colhead{V} & \colhead{Campaign} & \colhead{Notes} & \colhead{Class}\\ \colhead{} & \colhead{} & \colhead{Type} & \colhead{(mag)} & \colhead{} & \colhead{} & \colhead{ }}
\startdata
$\eta$ Tau & 200007767 & B7III & 2.986 & 4 & \tablenotemark{a} & SPB \\
27 Tau & 200007768 &  & 3.763 & 4 & \tablenotemark{a} & SPB \\
17 Tau & 200007769 & B6IIIe & 3.851 & 4 & \tablenotemark{a} & SPB \\
23 Tau & 200007770 & B6IVe & 4.305 & 4 & \tablenotemark{a} & SPB \\
20 Tau & 200007771 & B8III & 4.305 & 4 & \tablenotemark{a} & $\alpha^2\,\text{CVn}$ \\
19 Tau & 200007772 & B6IV & 4.448 & 4 & \tablenotemark{a} & SPB \\
28 Tau & 200007773 & B8Ve & 5.192 & 4 & \tablenotemark{a} & SPB \\
$\gamma$ Tau & 200007765 & G9.5III & 3.474 & 4 &  & RG \\
$\delta^{1}$ Tau & 200007766 & G9.5III & 3.585 & 4 &  & RG \\
$\alpha$ Vir & 212573842 & B1V & 0.97 & 6, 17 & Normal Mask & SPB \\
69 Vir & 212356048 & K0III & 4.75 & 6 &  & -- \\
$\zeta$ Sgr & 200062593 & A2.5V & 2.585 & 7 &  & $\gamma\,\text{Dor}$ \\
$\pi$ Sgr & 200062592 & F2II-III & 2.88 & 7 &  & Supergiant \\
$\tau$ Sgr & 200062591 & K1.5III & 3.31 & 7 &  & RG \\
$\xi^{2}$ Sgr & 200062590 & G8/K0II/III & 3.51 & 7 &  & RG \\
$o$ Sgr & 200062589 & G9III & 3.77 & 7 &  & RG \\
52 Sgr & 200062585 & B8/9V & 4.598 & 7 &  & SPB + Rotation \\
$\nu^{1}$ Sgr & 200062588 & K1II & 4.845 & 7 &  & -- \\
$\psi$ Sgr & 200062584 & K0/1III & 4.85 & 7 &  & -- \\
43 Sgr & 200062587 & G8II-III & 4.878 & 7 &  & -- \\
$\nu^{2}$ Sgr & 200062586 & K3-II-III & 4.98 & 7 &  & RG \\
$\epsilon$ Psc & 200068392 & G9IIIe & 4.28 & 8 &  & RG \\
$\zeta$ Psc A & 200068393 & A7IV & 5.187 & 8 &  & $\delta\,\text{Sct}$/$\gamma\,\text{Dor}$ \\
80 Psc & 200068394 & F2V & 5.5 & 8 &  & $\gamma\,\text{Dor}$ \\
42 Cet & 200068399 & G8IV & 5.87 & 8 &  & ? \\
33 Cet & 200068395 & K4/5III & 5.942 & 8 &  & -- \\
60 Psc & 200068396 & G8III & 5.961 & 8 &  & -- \\
73 Psc & 200068397 & K5III & 6.007 & 8 &  & -- \\
WW Psc & 200068398 & M2.5III & 6.14 & 8 &  & -- \\
HR 243 & 200068400 & G8/K0II/III & 6.368 & 8 &  & -- \\
HR 161 & 200068401 & K3III & 6.407 & 8 &  & -- \\
HR 6766 & 200069361 & G7:III & 4.56 & 9 &  & RG \\
HR 6842 & 200069360 & K3II & 4.627 & 9 &  & -- \\
4 Sgr & 200069357 & A0 & 4.724 & 9 &  & -- \\
11 Sgr & 200069358 & K0III & 4.98 & 9 &  & RG \\
7 Sgr & 200069362 & F2II-III & 5.34 & 9 &  & RG \\
15 Sgr & 200069359 & O9.7I & 5.37 & 9 &  & O \\
HR 6838 & 200069363 & K2III & 5.75 & 9 &  & -- \\
Y Sgr & 200069364 & F8II & 5.75 & 9 &  & Cepheid \\
HR 6716 & 200069365 & B0I & 5.77 & 9 &  & SPB \\
HR 6681 & 200069366 & A0V & 5.929 & 9 &  & -- \\
9 Sgr & 200069368 & O4V & 5.97 & 9 &  & Supergiant \\
16 Sgr & 200069367 & O9.5III & 6.02 & 9 &  & RG \\
HR 6825 & 200069369 & ApSip & 6.15 & 9 &  & $\gamma\,\text{Dor}$ \\
63 Oph & 200069370 & O8II & 6.2 & 9 &  & O \\
HR 6679 & 200069373 & A1V & 6.469 & 9 &  & -- \\
HD 165784 & 200069371 & A2I & 6.58 & 9 &  & -- \\
HD 161083 & 200069374 & F0V & 6.58 & 9 &  & $\delta\,\text{Sct}$/$\gamma\,\text{Dor}$ \\
5 Sgr & 200069372 & K0III & 6.64 & 9 &  & RG \\
HD 167576 & 200069378 & K1III & 6.66 & 9 &  & -- \\
HR 6773 & 200069380 & B3/5IV & 6.71 & 9 &  & EB + SPB \\
HD 163296 & 200071159 & A1Vpe & 6.85 & 9 &  & $\gamma\,\text{Dor}$ \\
HD 165052 & 200069379 & O6V+O8V & 6.87 & 9 &  & O \\
17 Sgr & 200069375 & G8/K0III & 6.886 & 9 &  & -- \\
HD 169966 & 200069376 & G8/K0III & 6.97 & 9 &  & -- \\
HD 162030 & 200069377 & K1III & 7.02 & 9 &  & -- \\
$\gamma$ Vir & 200084004 & F1V+F2Vm & 2.74 & 10 &  & $\gamma\,\text{Dor}$ \\
$\eta$ Vir & 200084005 & A2IV & 3.9 & 10 &  & $\delta\,\text{Sct}$ \\
21 Vir & 200084006 & B9V & 5.48 & 10 &  & -- \\
FW Vir & 200084007 & M3+IIICa0.5 & 5.71 & 10 &  & -- \\
HR 4837 & 200084008 & G8III & 5.918 & 10 &  & -- \\
HR 4591 & 200084009 & K1III & 6.316 & 10 &  & -- \\
HR 4613 & 200084010 & G8/K0III & 6.364 & 10 &  & -- \\
HD 107794 & 200084011 & K0III & 6.46 & 10 &  & -- \\
$\theta$ Oph & 200128906 & OB & 3.26 & 11 &  & $\beta$\,Cep \\
44 Oph & 200128907 & A3m & 4.153 & 11 &  & -- \\
45 Oph & 200128908 & F5III-IV & 4.269 & 11 &  & -- \\
51 Oph & 200128909 & A0V & 4.81 & 11 &  & Rotation \\
36 Oph & 200129035 & K2V+K1V & 5.03 & 11 &  & Rotation \\
$o$ Oph & 200128910 &  & 5.2 & 11 &  & ? \\
26 Oph & 200129034 & F3V & 5.731 & 11 &  & $\gamma\,\text{Dor}$ \\
HR 6472 & 200128911 & K0III & 5.83 & 11 &  & -- \\
HR 6366 & 200128913 & Fm & 5.911 & 11 &  & -- \\
HR 6365 & 200128912 & K0III & 5.977 & 11 &  & -- \\
191 Oph & 200128914 & K0III & 6.171 & 11 &  & RG \\
$\kappa$ Psc & 200164167 & A2Vp & 4.94 & 12 &  & Rotation + $\delta\,\text{Sct}$ \\
83 Aqr & 200164168 & F0V & 5.47 & 12 &  & $\delta\,\text{Sct}$/$\gamma\,\text{Dor}$ \\
24 Psc & 200164169 & K0II/III & 5.94 & 12 &  & -- \\
HR 8759 & 200164170 & G5II/III & 5.933 & 12 &  & RG \\
14 Psc & 200164171 & A2II & 5.87 & 12 &  & Supergiant \\
HR 8921 & 200164172 & K4/5III & 6.191 & 12 &  & -- \\
81 Aqr & 200164173 & K4III & 6.215 & 12 &  & RG \\
HR 8897 & 200164174 & K4III & 6.34 & 12 &  & -- \\
$\alpha$ Tau & 200173843 & K5+III & 0.86 & 13 & \tablenotemark{c} & -- \\
$\theta^{2}$ Tau & 200173845 & A7III & 3.41 & 13 & SC & $\delta\,\text{Sct}$ \\
$\epsilon$ Tau & 200173844 & G9.5III & 3.53 & 13 & \tablenotemark{d} & RG \\
$\theta^{1}$ Tau & 200173846 & G9IIIe & 3.84 & 13 &  & $^f$ \\
$\kappa^{1}$ Tau & 200173847 & A7IV & 4.201 & 13 & SC & $\delta\,\text{Sct}$ \\
$\delta^{3}$ Tau & 200173849 & A2IV & 4.25 & 13 & C4 & Supergiant \\
$\tau$ Tau & 200173850 & B3V & 4.258 & 13 &  & SPB \\
$\upsilon$ Tau & 200173848 & A8V & 4.282 & 13 & SC & $\delta\,\text{Sct}$ \\
$\rho$ Tau & 200173851 & A8V & 4.65 & 13 & SC & $\delta\,\text{Sct}$ \\
11 Ori & 200173853 & A1Vp & 4.661 & 13 &  & Rotation \\
HR 1427 & 200173855 & A6IV & 4.764 & 13 & SC & $\gamma\,\text{Dor}$? \\
15 Ori & 200173854 & F2IV & 4.82 & 13 &  & $\gamma\,\text{Dor}$ \\
75 Tau & 200173852 & K1III & 4.969 & 13 &  & RG \\
97 Tau & 200173857 & A7IV & 5.085 & 13 & SC & $\delta\,\text{Sct}$/$\gamma\,\text{Dor}$ \\
HR 1684 & 200173856 & K5III & 5.163 & 13 &  & -- \\
$\kappa^{2}$ Tau & 200173859 & F0V & 5.264 & 13 & SC & $\delta\,\text{Sct}$/$\gamma\,\text{Dor}$ \\
56 Tau & 200173861 & A0Vp & 5.346 & 13 &  & $\delta\,\text{Sct}$ \\
81 Tau & 200173860 & Am & 5.454 & 13 &  & -- \\
53 Tau & 200173864 & B9Vp & 5.482 & 13 &  & SPB \\
HR 1585 & 200173858 & K1III & 5.49 & 13 &  & RG \\
80 Tau & 200173866 & F0V & 5.552 & 13 &  & $\gamma\,\text{Dor}$ \\
51 Tau & 200173865 & F0V & 5.631 & 13 &  & $\delta\,\text{Sct}$ \\
HR 1403 & 200173867 & Am & 5.711 & 13 &  & -- \\
89 Tau & 200173868 & F0V & 5.776 & 13 &  & $\delta\,\text{Sct}$/$\gamma\,\text{Dor}$ \\
HR 1576 & 200173871 & B9V & 5.776 & 13 &  & SPB \\
98 Tau & 200173870 & A0V & 5.785 & 13 &  & EB + $\alpha^2\,\text{CVn}$ \\
99 Tau & 200173862 & K0III & 5.806 & 13 &  & RG \\
105 Tau & 200173869 & B2Ve & 5.92 & 13 &  & $\beta$\,Cep \\
HR 1554 & 200173874 & F2IV & 5.961 & 13 &  & $\delta\,\text{Sct}$/$\gamma\,\text{Dor}$ \\
HR 1385 & 200173875 & F4V & 5.965 & 13 & C4 & $\delta\,\text{Sct}$/$\gamma\,\text{Dor}$ \\
HR 1741 & 200173873 & K0III & 6.107 & 13 &  & -- \\
HR 1633 & 200173872 & K0 & 6.188 & 13 &  & RG \\
HR 1755 & 200173876 & K0III & 6.205 & 13 &  & RG \\
$\rho$ Leo & 200182931 & B1I & 3.87 & 14 & \tablenotemark{e} & Supergiant \\
58 Leo & 200182925 & K0.5IIIe & 4.838 & 14 &  & RG \\
48 Leo & 200182926 & G8.5IIIe & 5.07 & 14 &  & RG \\
53 Leo & 200182928 & A2V & 5.312 & 14 &  & $\delta\,\text{Sct}$ \\
65 Leo & 200182927 & K0III & 5.52 & 14 &  & RG \\
35 Sex & 200182929 & K1+K2III & 5.79 & 14 &  & RG \\
43 Leo & 200182930 & K3III & 6.08 & 14 &  & RG \\
$\delta$ Sco & 200194910 & B0.3IV & 2.32 & 15 &  & $\beta$\,Cep \\
$\gamma$ Lib & 200194911 & G8.5III & 3.91 & 15 &  & RG \\
$\iota^{1}$ Lib & 200194912 & B9IVp & 4.54 & 15 & \tablenotemark{b} & Rotation + SPB \\
41 Lib & 200194913 & G8III/IV & 5.359 & 15 &  & RG \\
$\zeta^{4}$ Lib & 200194914 & B3V & 5.499 & 15 &  & $\beta$\,Cep \\
HR 5762 & 200194915 & A2IV & 5.52 & 15 &  & -- \\
HR 5806 & 200194916 & K0III & 5.79 & 15 &  & RG \\
$\zeta^{3}$ Lib & 200194917 & K0III & 5.806 & 15 &  & RG \\
HR 5810 & 200194918 & K0III & 5.816 & 15 &  & RG \\
$\iota^{2}$ Lib & 200194919 & A2V & 6.066 & 15 & \tablenotemark{b} & $\delta\,\text{Sct}$ \\
HR 5620 & 200194920 & K0III & 6.14 & 15 &  & RG \\
28 Lib & 200194921 & G8II/III & 6.17 & 15 &  & RG \\
HD 138810 & 200194958 & K1III & 7.02 & 15 &  & -- \\
$\delta$ Cnc & 200200356 & K0+IIIb & 3.94 & 16 &  & -- \\
$\alpha$ Cnc & 200200357 & A5m & 4.249 & 16 &  & Rotation \\
$\xi$ Cnc & 200200358 & G8.5IIIe & 5.149 & 16 &  & -- \\
$o^{1}$ Cnc & 200200360 & A5III & 5.22 & 16 &  & -- \\
$\eta$ Cnc & 200200359 & K3III & 5.325 & 16, 18 &  & RG \\
45 Cnc & 200200728 & A3III+G7III & 5.65 & 16 & SC & $\delta\,\text{Sct}$ \\
$o^{2}$ Cnc & 200200361 & F0IV & 5.677 & 16 &  & -- \\
50 Cnc & 200200363 & A1Vp & 5.885 & 16, 18 &  & $\delta\,\text{Sct}$ \\
82 Vir & 200213053 & M1+III & 5.01 & 17 &  & -- \\
76 Vir & 200213054 & G8III & 5.21 & 17 &  & RG \\
68 Vir & 200213055 & K5III & 5.25 & 17 &  & -- \\
80 Vir & 200213056 & K0III & 5.706 & 17 &  & RG \\
HR 5106 & 200213057 & A0V & 5.932 & 17 &  & $\delta\,\text{Sct}$ \\
HR 5059 & 200213058 & A8V & 5.965 & 17 &  & $\gamma\,\text{Dor}$ \\
$\gamma$ Cnc & 200233186 & A1IV & 4.652 & 18 & C5 & -- \\
$\zeta$ Cnc & 200233643 & F8V+G0V & 4.67 & 18 & C5 & -- \\
60 Cnc & 200233188 & K5III & 5.44 & 18 & C5, C16 & -- \\
49 Cnc & 200233189 & A1Vp & 5.66 & 18 & C5 & Rotation + $\gamma\,\text{Dor}$ \\
HR 3264 & 200233190 & K1III & 5.798 & 18 & C5 & RG \\
29 Cnc & 200233192 & A5V & 5.948 & 18 & C5 & $\delta\,\text{Sct}$/$\gamma\,\text{Dor}$ \\
HR 3222 & 200233193 & G8III & 6.047 & 18 & C5 & -- \\
21 Cnc & 200233196 & M2III & 6.08 & 18 & C5 & -- \\
25 Cnc & 200233644 & F5IIIm? & 6.1 & 18 & C5 & -- \\
HR 3558 & 200233195 & K1III & 6.146 & 18 & C5 & -- \\
HR 3541 & 200233194 & C-N4.5 & 6.4 & 18 & C5 & --
\enddata
\tablerefs{$^a$:~\citet{White2017}; $^b$:~\citet{Buysschaert2018}; $^c$:~\citet{Farr2018}; $^d$:~\citet{Arentoft2019}; $^e$:~\citet{Aerts2018}; $^f$:~Light curve shows RG pulsations, but is also significantly contaminated by the higher amplitude $\delta$~Sct pulsations of the nearby $\theta^2$~Tau.}
\tablecomments{Some targets are known by proper names. $\eta$~Tau: Alcyone; 27~Tau: Atlas; 17~Tau: Electra; 20~Tau: Maia; 23~Tau: Merope; 19~Tau: Taygeta; 28~Tau: Pleione; $\zeta$~Sgr: Ascella; $\pi$~Sgr: Albaldah; $\nu^{1}$~Sgr: Ainalrami; $\zeta$~Psc~A: Revati; $\gamma$~Vir: Porrima; $\eta$~Vir: Zaniah; $\alpha$~Tau: Aldebaran; $\delta$~Sco: Dschubba; $\gamma$~Lib: Zubenelhakrabi; $\delta$~Cnc: Asellus Australis; $\alpha$~Cnc: Acubens; $\alpha$~Vir: Spica; 36~Oph: Guniibuu; $\gamma$~Tau: Prima Hyadum; $\delta^{1}$~Tau: Secunda Hyadum; $\theta^{2}$~Tau: Chamukuy; $\epsilon$~Tau: Ain; $\xi$~Cnc: Nahn; $\gamma$~Cnc: Asellus Borealis; $\zeta$~Cnc: Tegmine}
\end{deluxetable}

\begin{deluxetable}{cccc}
\tablecaption{Global asteroseismic parameters for the 31 red giants for which solar-like oscillations were detected.\label{rgs}}
\tablehead{\colhead{Name} & \colhead{EPIC} & \colhead{\numax} & \colhead{\Dnu}\\ \colhead{} & \colhead{} & \colhead{(\muHz)} & \colhead{(\muHz)}}
\startdata
$\gamma$ Tau & 200007765 & 62.89 $\pm$ 1.44 & 5.56 $\pm$ 0.17 \\
$\delta^{1}$ Tau & 200007766 & 62.59 $\pm$ 1.74 & 5.72 $\pm$ 0.07 \\
$\nu^{2}$ Sgr & 200062586 & 7.29 $\pm$ 0.15 & 1.31 $\pm$ 0.05 \\
$o$ Sgr & 200062589 & 46.28 $\pm$ 1.02 & 4.82 $\pm$ 0.06 \\
$\xi^{2}$ Sgr & 200062590 & 11.71 $\pm$ 0.65 & 1.87 $\pm$ 0.15 \\
$\tau$ Sgr & 200062591 & 19.85 $\pm$ 0.80 & 2.46 $\pm$ 0.07 \\
$\pi$ Sgr & 200062592 & 46.95 $\pm$ 0.43 & 5.97 $\pm$ 0.20 \\
$\epsilon$ Psc & 200068392 & 33.31 $\pm$ 1.22 & 3.62 $\pm$ 0.07 \\
11 Sgr & 200069358 & 38.03 $\pm$ 0.84 & 4.01 $\pm$ 0.13 \\
HR 6766 & 200069361 & 20.60 $\pm$ 4.19 & 2.42 $\pm$ 0.41 \\
7 Sgr & 200069362 & 13.59 $\pm$ 0.97 & 1.98 $\pm$ 0.20 \\
HR 6716 & 200069365 & 10.68 $\pm$ 3.38 & 1.77 $\pm$ 0.28 \\
16 Sgr & 200069367 & 13.76 $\pm$ 0.34 & 2.23 $\pm$ 0.11 \\
5 Sgr & 200069372 & 47.78 $\pm$ 0.95 & 4.65 $\pm$ 0.05 \\
191 Oph & 200128914 & 29.19 $\pm$ 0.92 & 3.91 $\pm$ 0.10 \\
HR 8759 & 200164170 & 10.14 $\pm$ 0.39 & 1.56 $\pm$ 0.05 \\
81 Aqr & 200164173 & 11.38 $\pm$ 0.23 & 1.69 $\pm$ 0.06 \\
$\epsilon$ Tau & 200173844 & 54.46 $\pm$ 1.44 & 5.13 $\pm$ 0.13 \\
75 Tau & 200173852 & 34.95 $\pm$ 0.96 & 4.15 $\pm$ 0.04 \\
HR 1585 & 200173858 & 9.38 $\pm$ 1.01 & 1.48 $\pm$ 0.10 \\
99 Tau & 200173862 & 21.44 $\pm$ 1.07 & 2.41 $\pm$ 0.07 \\
HR 1755 & 200173876 & 18.78 $\pm$ 0.41 & 2.04 $\pm$ 0.04 \\
58 Leo & 200182925 & 17.01 $\pm$ 0.46 & 1.97 $\pm$ 0.23 \\
48 Leo & 200182926 & 53.32 $\pm$ 0.79 & 5.43 $\pm$ 0.04 \\
65 Leo & 200182927 & 61.65 $\pm$ 1.38 & 6.43 $\pm$ 0.03 \\
35 Sex & 200182929 & 11.52 $\pm$ 0.15 & 1.52 $\pm$ 0.05 \\
43 Leo & 200182930 & 71.61 $\pm$ 2.81 & 7.20 $\pm$ 0.08 \\
$\gamma$ Lib & 200194911 & 34.89 $\pm$ 0.98 & 3.57 $\pm$ 0.10 \\
41 Lib & 200194913 & 54.25 $\pm$ 1.79 & 5.19 $\pm$ 0.03 \\
HR 5806 & 200194916 & 53.22 $\pm$ 0.75 & 4.91 $\pm$ 0.06 \\
$\zeta^{3}$ Lib & 200194917 & 44.18 $\pm$ 1.00 & 3.55 $\pm$ 0.26 \\
HR 5810 & 200194918 & 45.02 $\pm$ 0.46 & 4.46 $\pm$ 0.03 \\
HR 5620 & 200194920 & 96.84 $\pm$ 0.74 & 9.28 $\pm$ 0.03 \\
28 Lib & 200194921 & 41.05 $\pm$ 0.86 & 4.10 $\pm$ 0.17 \\
$\eta$ Cnc & 200200359 & 22.91 $\pm$ 0.86 & 2.65 $\pm$ 0.03 \\
76 Vir & 200213054 & 40.02 $\pm$ 2.62 & 3.76 $\pm$ 0.09 \\
80 Vir & 200213056 & 36.98 $\pm$ 1.83 & 4.38 $\pm$ 0.08 \\
HR 3264 & 200233190 & 22.93 $\pm$ 0.17 & 3.00 $\pm$ 0.18
\enddata
\end{deluxetable}



\end{document}